\documentclass[12pt]{iopart}
\usepackage{txfonts}
\usepackage{iopams}
\usepackage{graphicx}

\begin{document}

\title{Direct scheme for measuring the geometric quantum discord}
\author{Jia-sen Jin, Feng-yang Zhang, Chang-shui Yu, and He-shan Song}
\address{School of Physics and Optoelectronic Technology, Dalian University of Technology, Dalian 116024 China}
\ead{\mailto{ycs@dlut.edu.cn},\mailto{hssong@dlut.edu.cn}}

\begin{abstract}
We propose a scheme to directly measure the exact value of geometric quantum discord of an arbitrary two-qubit state. We only need to perform the projective measurement in the all anti-symmetric subspace and our scheme is parametrically efficient in contrast to the widely adopted quantum state tomography scheme in the sense of less parameter estimations and projectors. Moreover, the present scheme can be easily realized with the current experimental techniques.
\end{abstract}

\pacs{03.65.Ta, 03.65.Ud}
\maketitle

\section{Introduction}
Quantum correlation is believed to be a unique property possessed by a quantum system and plays an essential role in quantum information theory. Traditionally, studies on quantum correlation have mainly focused on entanglement \cite{en}, which is considered to be a crucial resource for the vast majority of quantum information processing. However, not all advantages of quantum information tasks attribute to quantum entanglement. For example, the deterministic quantum computation with one qubit model is believed to estimate the trace of a unitary matrix more efficiently than any classical information processors, whereas there is little or no entanglement during the whole processing \cite{DQC1}. This implies the possible presence of other type of, beyond quantum entanglement, quantum correlation measured by quantum discord in separable (unentangled) state \cite{QD1,QD2}.

Even though quantum discord has attracted considerable attention in quantum information theory involving quantum computation \cite{QC0,QC1,QC2}, dynamics of quantum discord \cite{dyD1,dyD2,dyD3}, operational interpretations of quantum discord \cite{m1,m2,d1,d2}, etc, only geometric quantum discord (GQD) defined by the distance between a general quantum state and its closest classical state is analytically calculative in mathematics \cite{GQD}. However, can we directly measure the GQD in experiment? It seems not to be an easy work, since the quantum discord is a complicated nonlinear function of the density matrix of the state. To the best of our knowledge, the only method used to exactly estimate the value of quantum discord in laboratory is the so-called quantum state tomography (QST) \cite{xujs1,xujs2,tomo3}, where one has to measure a complete set of observables to reconstruct the density matrix, and then evaluates the exact value of quantum discord by mathematical measure. It is obvious that the number of observables will grow exponentially with the increasing dimension. Another more direct method is the witness of quantum discord, but unfortunately it requires some \emph{priori} knowledge on the state to be detected \cite{witness1} and only provides a lower bound of quantum discord \cite{witness2}.

In this paper, we propose an experimental scheme to directly measure the geometric quantum discord of a completely unknown two-qubit state. The present scheme can be used to measure the exact value of GQD rather than its lower bound. In addition, the GQD is measured by estimation of three parameters in joint measurements of a small number (not more than six) of copies. Compared with QST, our scheme has the following advantages: (i) our scheme requires less projective measurements than QST; (ii) in our scheme we only need to estimate 3 parameters instead of 15 parameters as done in QST; (iii) the GQD can be obtained with only local projective measurements on subsystems with the assistance of classical communications.

The paper is organized as follows. In Section II, we firstly give a brief review of the GQD, and then we demonstrate the scheme for directly measuring of GQD of a two-qubit state in detail. The conclusion is drawn finally.

\section{Scheme for directly measuring of geometric quantum discord}
\subsection{Geometric quantum discord}
In Ref. \cite{GQD}, Daki$\mathrm{\acute{c}}$ \emph{et al.} introduced the GQD to measure the quantum correlation from the geometric perspective, which has the following form,
\begin{equation}
D(\rho)=\mathrm{min}_{\chi\in\Omega_0}||\rho-\chi||^2,
\end{equation}
where $\Omega_0$ denotes the set of classical (zero-discord ) states and $||X-Y||^2=\mathrm{tr}(X-Y)^2$ stands for the square norm in the Hilbert-Schmidt space. In particular, it is possible to obtain an explicit expression of GQD for a general two-qubit state \cite{GQD}. In the Bloch representation, a two-qubit state $\rho$ can be written as follows,
\begin{equation}
\rho=\frac{1}{4}(\mathbb{I}\otimes \mathbb{I}+\sum_{i=1}^3x_i\sigma^i\otimes \mathbb{I}+\sum_{i=1}^3y_i\mathbb{I}\otimes\sigma^i+\sum_{i,j=1}^3t_{ij}\sigma^i\otimes\sigma^j),
\end{equation}
where $\mathbb{I}$ is the identity matrix, $\sigma^i (i=1,2,3)$ are the Pauli matrices, $x_i=\mathrm{tr}(\sigma^i\otimes\mathbb{I})\rho$, $y_i=\mathrm{tr}(\mathbb{I}\otimes\sigma^i)\rho$ are the components of the local Bloch vectors $\vec{x}$ and $\vec{y}$, respectively, and $t_{ij}=\mathrm{tr}(\sigma^i\otimes\sigma^j)\rho$ are components of correlation matrix $T$. The classical state in $\Omega_0$ is of the form $\chi=p_1|\psi_1\rangle\langle\psi_1|\otimes\rho_1+p_2|\psi_2\rangle\langle\psi_2|\otimes\rho_2$ with $p_1+p_2=1$, then the GQD of $\rho$ is given as
\begin{equation}
D(\rho)=\frac{1}{4}(||\vec{x}||^2+||T||^2-\lambda_{\mathrm{max}}),\label{D}
\end{equation}
with $\lambda_{\mathrm{max}}$ being the largest eigenvalue of the matrix $K=\vec{x}\vec{x}^t+TT^t$ and $||T||^2=\mathrm{tr}(TT^t)$. The superscript $t$ denotes the transpose of vectors or matrices.

The GQD presented in Eq. (\ref{D}) can be rewritten as follows,
\begin{equation}
D(\rho)=\frac{1}{4}(\sum_{i=1}^3\lambda_i-\lambda_{\mathrm{max}}),\label{D1}
\end{equation}
where $\lambda_i$'s are the eigenvalues of matrix $K$. In the derivation of Eq. (\ref{D1}), we have adopted the facts $||\vec{x}||^2=\mathrm{tr}(\vec{x}\vec{x}^t)$ and $||T||^2=\mathrm{tr}(TT^t)$.

Recall that the GQD is not a symmetric measure under the permutation of the subsystems, if we choose the zero-discord state in the form of $\chi=p_1\rho_1\otimes|\psi_1\rangle\langle\psi_1|+p_2\rho_2\otimes|\psi_2\rangle\langle\psi_2|$, then the GQD of $\rho$ under a permutation of the subsystems is given as
\begin{equation}
D'(\rho)=\frac{1}{4}(\sum_{i=1}^3\lambda'_i-\lambda'_{\mathrm{max}}),\label{Dp}
\end{equation}
with $\lambda'_i$ being the eigenvalue of matrix $K'=\vec{y}\vec{y}^t+T^tT$. Without loss of generality, we will mainly focus on the measurement of $D(\rho)$ in the following discussions.

\subsection{Directly measuring the GQD}

In this subsection, we will demonstrate our scheme explicitly. From Eq. (\ref{D1}) we can see that the acquirement of the eigenvalues of matrix $K$ is sufficient to estimate the GQD of $\rho$, this can be achieved if we can obtain the moments $M_k=\sum_{i=1}^3(\lambda_i)^k$, $k=1,2,3$ \cite{horo1}. Although it is known that nonlinear functions of the state parameters cannot be directly obtained by performing measurement on a single copy, we can try to estimate them with several copies. In fact, such an approach has been used in detecting the entanglement of an unknown quantum state \cite{c1,c2,c3}. In our scheme, in order to obtain each $M_k$ we need $2k$ copies of states at each steps. For conciseness, we employ $\rho_{a_mb_m}$ to denote the $m$th copy and $a$,$b$ to denote the subsystems of each copy, hereinafter.

Let us start with the attainment of $M_1$, in this procedure we need 2 copies for each projective measurement. Because $M_1$ is defined as the summation of the eigenvalues of $K$ which mathematically equals to the trace of $K$, we have $M_1=\mathrm{tr}(\vec{x}\vec{x}^t+TT^t)$. Taking into accounts the expressions of $\vec{x}$ and $T$, $M_1$ can be written as follows,
\begin{eqnarray}
M_1&=&\mathrm{tr}(\vec{x}\vec{x}^t+TT^t)=\mathrm{tr}(\vec{x}\vec{x}^t)+\mathrm{tr}(TT^t)\cr\cr&=&
\sum_{i=1}^3[\mathrm{tr}(\sigma^i_{a_1}\otimes \mathbb{I}_{b_1})\rho_{a_1b_1}\cdot\mathrm{tr}(\sigma^i_{a_2}\otimes \mathbb{I}_{b_2})\rho_{a_2b_2}]\cr\cr
&&+\sum_{i,j=1}^3[\mathrm{tr}(\sigma^i_{a_1}\otimes \sigma^j_{b_1})\rho_{a_1b_1}\cdot\mathrm{tr}(\sigma^i_{a_2}\otimes \sigma^j_{b_2})\rho_{a_2b_2}]\cr\cr
&=&\mathrm{tr}\sum_{i=1}^3(\sigma^i_{a_1}\otimes\sigma^i_{a_2}\otimes\mathbb{I}_{b_1}\otimes\mathbb{I}_{b_2})(\rho_{a_1b_1}\otimes\rho_{a_2b_2})\cr\cr
&&+\mathrm{tr}\sum_{i,j=1}^3(\sigma^i_{a_1}\otimes\sigma^i_{a_2}\otimes\sigma^j_{b_1}\otimes\sigma^j_{b_2})(\rho_{a_1b_1}\otimes\rho_{a_2b_2})\cr\cr
&=&\mathrm{tr}[(U_{a_1a_2}\otimes V_{b_1b_2})(\rho_{a_1b_1}\otimes\rho_{a_2b_2})],\label{m1}
\end{eqnarray}
In the last equal sign of Eq. (\ref{m1}), we have defined
\begin{equation}
U_{a_ma_n}=\sum_{i=1}^3\sigma^i_{a_m}\otimes\sigma^i_{a_n}=-4P^-_{a_ma_n}+\mathbb{I}_{a_ma_n},\label{U}
\end{equation}
\begin{equation}
V_{b_mb_n}=\mathbb{I}_{b_m}\otimes\mathbb{I}_{b_n}+\sum_{i=1}^3\sigma^i_{b_m}\otimes\sigma^i_{b_n}=-4P^-_{b_mb_n}+2\mathbb{I}_{b_mb_n},\label{V}
\end{equation}
where $P^-_{s_ms_n}=|\Psi^-\rangle_{s_ms_n}\langle\Psi^-|$ and $|\Psi^-\rangle_{s_ms_n}=(|0\rangle_{s_m}|1\rangle_{s_n}-|1\rangle_{s_m}|0\rangle_{s_n})/\sqrt{2}$ with $s=a,b$ and here $m,n=1,2$. It is obvious that the operators $U_{a_ma_n}$ and $V_{b_mb_n}$ are local two-qubit operators that performed on subsystems $a$ and $b$, respectively.

Similarly, we can represent $M_2$ and $M_3$ in terms of the expectation values of the tensor products of $U_{a_ma_n}$ and $V_{b_mb_n}$ with 4 and 6 copies of states, respectively. The results are given as follows,
\begin{eqnarray}
M_2&=&\mathrm{tr}(\vec{x}\vec{x}^t+TT^t)^2\cr\cr
&=&\mathrm{tr}[(U_{a_1a_4}\otimes U_{a_2a_3}\otimes V_{b_1b_2}\otimes V_{b_3b_4})\cr\cr
&&\times(\rho_{a_1b_1}\otimes\rho_{a_2b_2}\otimes\rho_{a_3b_3}\otimes\rho_{a_4b_4})],\label{m2}
\end{eqnarray}
\begin{eqnarray}
M_3&=&\mathrm{tr}(\vec{x}\vec{x}^t+TT^t)^3\cr\cr
&=&\mathrm{tr}[(U_{a_1a_6}\otimes U_{a_2a_3}\otimes U_{a_4a_5}\otimes V_{b_1b_2}\otimes V_{b_3b_4}\otimes V_{b_5b_6})\cr\cr
&&\times(\rho_{a_1b_1}\otimes\rho_{a_2b_2}\otimes\rho_{a_3b_3}\otimes\rho_{a_4b_4}\otimes\rho_{a_5b_5}\otimes\rho_{a_6b_6})].\label{m3}
\end{eqnarray}

\begin{figure}[tbp]
\includegraphics[width=1\columnwidth]{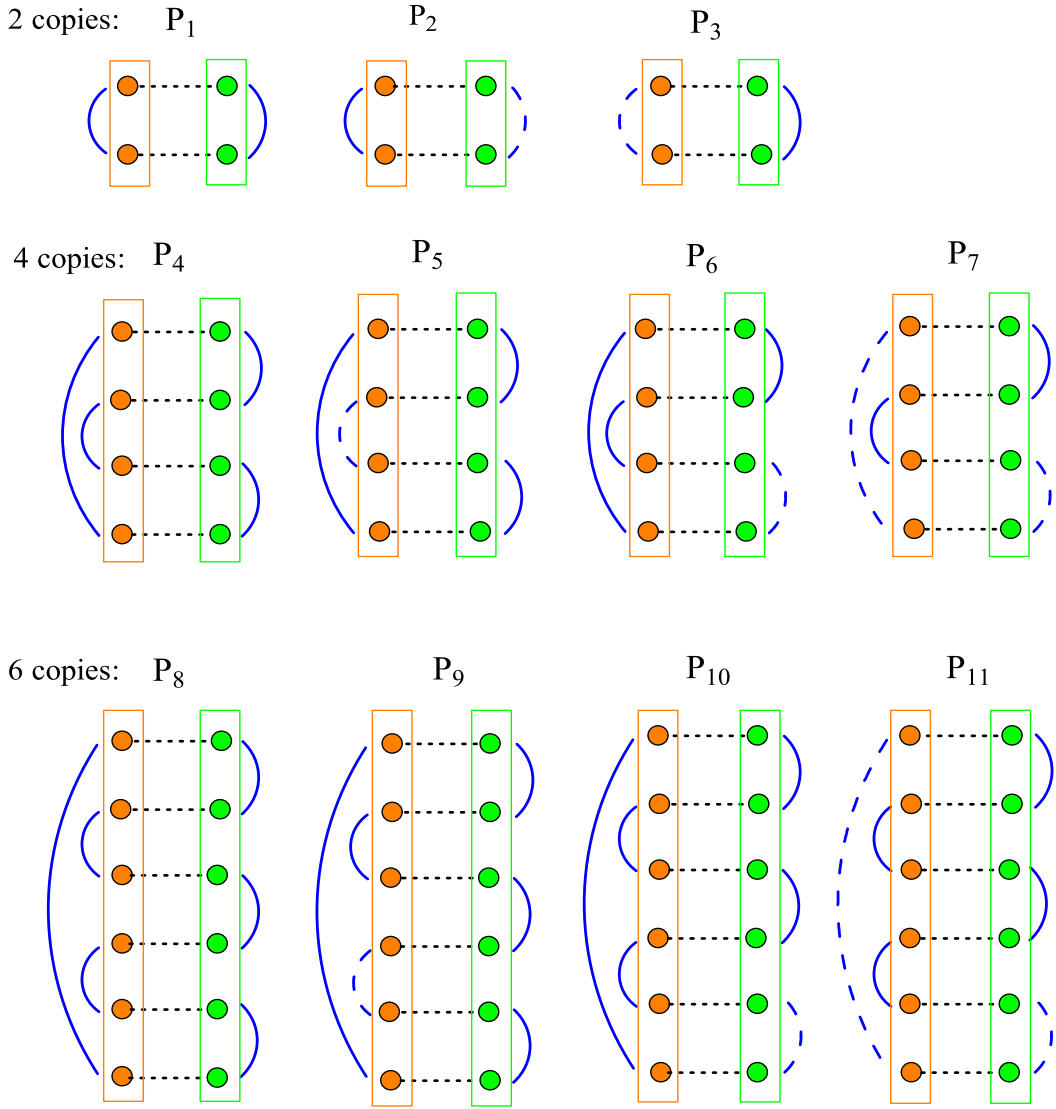}
\caption{(Color online) Schematic diagram of the required projective measurements. The circles in orange (green) boxes denote the qubits of subsystem a (b). In each subplot, a pair of orange and green circles (connected by a black dotted line) at the $m$th row denotes the $i$th copy of state $\rho_{a_mb_m}$. The blue solid curve connecting two circles denotes the operator $P^-$ and the blue dashed curve denotes the identity operator $\mathbb{I}$. For example, the last subplot demonstrates the projective measurements $P_{11}=\mathbb{I}_{a_1a_6}\otimes P^-_{a_2a_3}\otimes P^-_{a_4a_5}\otimes P^-_{b_1b_2}\otimes P^-_{b_3b_4}\otimes \mathbb{I}_{b_5b_6}$.
} \label{1}
\end{figure}

By substituting Eqs. (\ref{U}) and (\ref{V}) to Eqs. (\ref{m1}), (\ref{m2}), and (\ref{m3}), we can represent each $M_k$ in terms of the sum of the tensor products of $P_{s_ms_n}$, each term in the summation corresponds to a joint measurement on the multiple copies. It is interesting that the joint measurements are composed of a number of projective measurement which are performed on two qubits in the same subsystem, that means the value of $M_k$ can be obtained by performing a set of local two-qubit projective measurements. Although the new expressions of $M_k$'s seem to be complex, even tedious, they can be simplified since the outcomes of the joint measurements are not independent. After simple calculations, we find that only $11$ projective measurements are required. The number of the projective measurements is less than that required in QST. The $11$ required projectors are given as follows, see also Fig. \ref{1},
for 2 copies case:
\begin{eqnarray}
P_1&=&P^-_{a_1a_2}\otimes P^-_{b_1b_2},\cr\cr
P_2&=&P^-_{a_1a_2}\otimes \mathbb{I}_{b_1b_2},\cr\cr
P_3&=&\mathbb{I}_{a_1a_2}\otimes P^-_{b_1b_2},\label{p1}
\end{eqnarray}
for 4 copies case:
\begin{eqnarray}
P_4&=&P^-_{a_1a_4}\otimes P^-_{a_2a_3}\otimes P^-_{b_1b_2}\otimes P^-_{b_3b_4},\cr\cr
P_5&=&P^-_{a_1a_4}\otimes \mathbb{I}_{a_2a_3}\otimes P^-_{b_1b_2}\otimes P^-_{b_3b_4},\cr\cr
P_6&=&P^-_{a_1a_4}\otimes P^-_{a_2a_3}\otimes P^-_{b_1b_2}\otimes \mathbb{I}_{b_3b_4},\cr\cr
P_7&=&\mathbb{I}_{a_1a_4}\otimes P^-_{a_2a_3}\otimes P^-_{b_1b_2}\otimes \mathbb{I}_{b_3b_4},\label{p2}
\end{eqnarray}
for 6 copies case:
\begin{eqnarray}
P_8&=&P^-_{a_1a_6}\otimes P^-_{a_2a_3}\otimes P^-_{a_4a_5}\otimes P^-_{b_1b_2}\otimes P^-_{b_3b_4}\otimes P^-_{b_5b_6},\cr\cr
P_9&=&P^-_{a_1a_6}\otimes P^-_{a_2a_3}\otimes \mathbb{I}_{a_4a_5}\otimes P^-_{b_1b_2}\otimes P^-_{b_3b_4}\otimes P^-_{b_5b_6},\cr\cr
P_{10}&=&P^-_{a_1a_6}\otimes P^-_{a_2a_3}\otimes P^-_{a_4a_5}\otimes P^-_{b_1b_2}\otimes P^-_{b_3b_4}\otimes \mathbb{I}_{b_5b_6},\cr\cr
P_{11}&=&\mathbb{I}_{a_1a_6}\otimes P^-_{a_2a_3}\otimes P^-_{a_4a_5}\otimes P^-_{b_1b_2}\otimes P^-_{b_3b_4}\otimes \mathbb{I}_{b_5b_6},\label{p3}
\end{eqnarray}

Set the outcome of projective measurement $P_i$ is $c_i$, then the values of $M_k$ are given as follows,
\begin{eqnarray}
M_1&=&16c_1-8c_2-4c_3+2,\cr\cr
M_2&=&256c_4+128c_7-128(c_5+2c_6)-16(c_3+2c_2)\cr\cr&&+16(c_3^2+4c_2^2)+4,\cr\cr
M_3&=&4096c_8-16(32c_2^3+4c_3^3-24c_2^2-6c_3^2+6c_2-3c_3)\cr\cr
&&+192(8c_7^2+16c_2c_6+4c_3c_5-8c_2c_7-4c_3c_7+c_2c_3)\cr\cr
&&+384(c_7+8c_{11}-c_5-2c_6-8c_9-16c_{10})+8,\label{m123}
\end{eqnarray}
Thus, we have obtained the moments $M_k$'s with direct projective measurements, the eigenvalues $\lambda_i$'s can be sequentially determined from the moments using the techniques of Ref. \cite{spec}. Note that we can also obtain the moments of $K'$ by interchanging $c_2$, $c_5$, and $c_9$ with $c_3$, $c_6$, and $c_{10}$, respectively, in Eq. (\ref{m123}).

In the analysis above, we find that in principle the GQD can be directly and locally measured in experiment. This will be valuable for the case of the correlated qubits are shared by two distant participants. Moreover, different from the previous works on entanglement detection, which requires projective measurements on multiple-qubit space in certain steps \cite{cai}, the measurements involved in our scheme are all two-qubit projective measurements, which is more feasible in experiments, for instance, the two-qubit projective measurements have already been used for detection of lower bounds of entanglement in photonic systems \cite{exp1,exp2}. Moreover, we have also roughly compared our scheme with QST utilizing the criterion mentioned in Ref. \cite{horo2}. In our scheme, the number of the parameters to be measured, $r_p$, is $3$, which is less than that in QST, $15$. In this sense we can say that our scheme is parametrically efficient. When the number of copies ,$r_c$, is concerned, our scheme is a little more consumptive than QST, that is, in each ``round'' of our scheme the number of consumed copies is larger than that in QST ($44>15$). However, the factor $r=r_pr_c$ is here ($r=132$) less than for the QST schemes ($r=225$).

Finally, we give a brief discussion on the experimental implementation of our scheme. As an example, we restrict our discussions in photonic system, which has mature manipulation techniques and widely applications in quantum information processing. The required two-qubit projective measurements can be easily realized with the help of beam splitters and single-photon detectors. There also exists two challenges in the experimental implementation: the preparation of the identical copies of an unknown state and the simultaneity of the performance of joint projective measurements. We adopt the method mentioned in Ref. \cite{exp2} to overcome these two points. The correlated photon pairs sequentially generated by the same photon source (undergo the same decoherence channel) are sent into different optical pathes, and by carefully arranging the lengthes of these optical pathes the photons belong to the same subsystem but generated at different times will arrive at the measuring device simultaneously. As a consequence, both the requirements of the identity of copies and the coincidence of measurements are satisfied. In addition, since the measurement settings for $P_1$, $P_4$, and $P_8$ can also be used to perform the other $8$ projective measurements, we need only three measurement settings.

\section{Conclusions}

In conclusion, we have shown that the exact GQD for an arbitrary unknown two-qubit state instead of a lower bound can be directly and locally measured in experiment, provided that not more than six copies of states to be measured are available. Furthermore, our scheme is parametrically efficient than the widely adopted QST scheme since we need measure only 3 parameters, \emph{i.e.}, three moments of matrix $K$, to obtain the quantum discord. Moreover, the attainment of quantum discord can be achieved by local projective measurements and classical communications, the number of projective measurements is less than that required in QST. In particular, these measurements are all projected onto two-qubit space which can be easily implemented in experiments. We also expect that the basic idea of our scheme may provide a new point of view to understand the quantum correlation.

\ack{This work was supported by the National Natural Science Foundation of China, under Grants No. 10805007 and No. 11175033, and the Doctoral Startup Foundation of Liaoning Province.}

\end{document}